\documentclass[twoside]{photon2007}
\usepackage[latin1]{inputenc}
\usepackage[dvips]{graphicx,epsfig,color}
\usepackage{wrapfig,rotating}
\usepackage{amssymb,amsmath,array}

\def\gsim{\raise0.3ex\hbox{$>$\kern-0.75em\raise-1.1ex\hbox{$\sim$}}}
\def\lsim{\raise0.3ex\hbox{$<$\kern-0.75em\raise-1.1ex\hbox{$\sim$}}}
\def\zp#1#2#3{{ Z. Phys. }{\bf #1}(19#2)#3}
\def\pr#1#2#3{{ Phys. Rev. }{\bf #1}(19#2)#3}

\begin{document}
\title{Prompt photon production in photoproduction, DIS and hadronic collisions} 
\author{Gudrun Heinrich\thanks{Invited talk given at the International Con\-ference Photon2007, Paris, July 9-13, 2007\,\cite{url}.}
\vspace{.3cm}\\
School of Physics, The University of Edinburgh, Edinburgh EH9 3JZ,
Scotland, UK \\
\vspace{.1cm}\\
}

\maketitle

\begin{abstract}
Recent results on prompt photon production in photoproduction, deeply inelastic 
scattering  and hadronic collisions 
are reviewed and the importance of photons for LHC experiments is 
briefly discussed. 
\end{abstract}

\section{Introduction}

Photons have always played a very important role in particle physics,  historically to 
develop the quantum theory of electro\-dyna\-mics (QED) 
and experimentally due to their omnipresence and 
clean signature. Another interesting aspect is
their ``dual nature" with respect to the strong interactions: 
On one hand, the photon acts like a pointlike particle described by QED, 
on the other hand, photons also have a ``hadronic face":
energetic partons can  ``fragment" into a large-$p_T$ photon and hadronic energy, 
this process being described by photon fragmentation
functions\,\cite{Gluck:1992zx,Bourhis:2000gs}.  
Similarly, initial state photons also can be ``resolved" 
into their hadronic structure 
by a hard interaction, leading to the concept of 
photon structure functions, 
see e.g. \cite{Krawczyk:2000mf,Aurenche:2005da} for recent reviews. 
Making use of this dual nature, reactions invol\-ving photons are an ideal tool to study 
QCD in various aspects (for recent literature, see e.g. \cite{Ajduk:2006gb,Klasen:2002xb}). 

Furthermore, final state photons play an important role in the search for a Higgs 
boson  with mass below $\sim$ 140 GeV, 
where the decay into two photons is a very prominent channel. 
In addition, they are important signatures for various 
scenarios of Physics Beyond the Standard Model, 
as they appear e.g. in  the decay  
chain of SUSY particles or Kaluza-Klein excitations.

Therefore it is very important to understand the Standard Model physics involving photons. 
For that matter we can learn a lot from past and present experiments. 
In the following, we will highlight some of the results obtained recently 
involving large-$p_T$ (``prompt") photons in the  final state.
Due to the limited scope of this article, we only cover HERA and hadron collider 
kinematics.

\section{Photoproduction of prompt photons}

High energy electron-proton scattering, as it has been carried out 
at the DESY $ep$ collider HERA,  
is dominated by so-called photoproduction processes, where the electron 
is scattered at small angles, emitting a quasi-real photon which 
scatters with the proton. The spectrum of these photons can be described by the 
Weizs\"acker-Williams approxi\-mation\,\cite{WW,Frixione:1993yw}. 
The $\gamma-p$ scattering processes are of special interest 
since they are sensitive to both the partonic structure of the photon 
as well as of the proton. 
As will be explained below, they offer the possibility  to constrain the 
(presently poorly known) gluon distributions in the photon, 
since in a certain kinematic region the subprocess 
$q g\to \gamma q$, where the gluon is stemming from a resolved photon,
is dominating\,\cite{Fontannaz:2003yn}. 

The cross section for $e p\to\gamma X$ can symbo\-li\-cally 
be written as a convolution of the parton densities for the incident particles
(resp. fragmentation function for an outgoing parton  fragmenting into a
photon) with the partonic cross section $\hat \sigma$:
\begin{eqnarray}
d\sigma^{ep\to\gamma X}(P_p,P_e,P_{\gamma})=
\sum_{a,b,c}\int dx_e\int d x_p\nonumber\\
\int
dz\, F_{a/e}(x_e,M)F_{b/p}(x_p,M_p)D_{\gamma/c}(z,M_F)\nonumber\\
d\hat\sigma^{ab\to c
X}(x_pP_p,x_eP_e,P_{\gamma}/z,\mu,M,M_p,M_F)\;,\nonumber\\
\hspace*{-1cm}\label{dsigma}
\end{eqnarray}
where $M,M_p$ are the initial state factorisation scales, $M_F$ the 
final state factorisation scale, $\mu$ the
renormalisation scale and $a,b,c$ run over parton types.

\begin{figure}[htb]
\begin{picture}(100,210)(0,-120)
\put(0,0){\epsfig{file=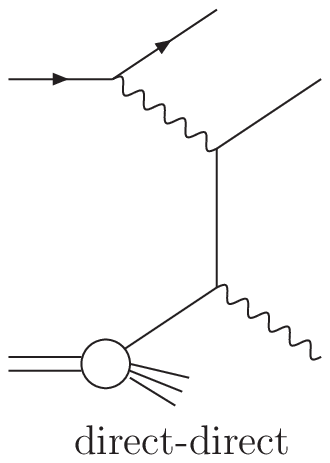,height=3.2cm}}
\put(100,0){\epsfig{file=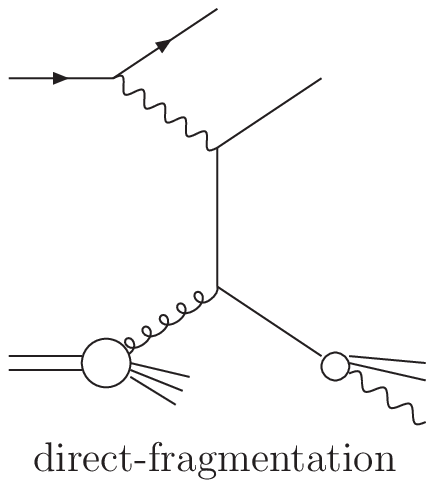,height=3.2cm}}
\put(0,-120){\epsfig{file=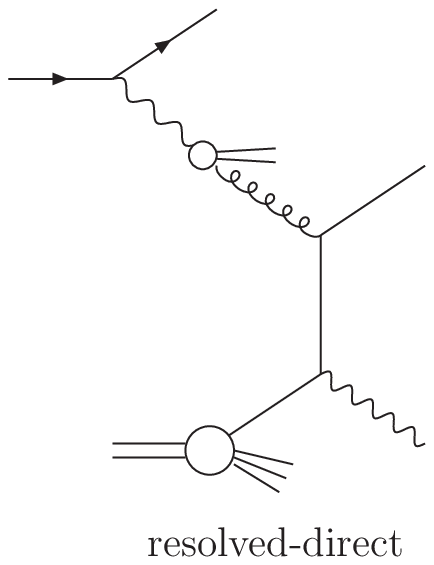,height=3.8cm}}
\put(100,-120){\epsfig{file=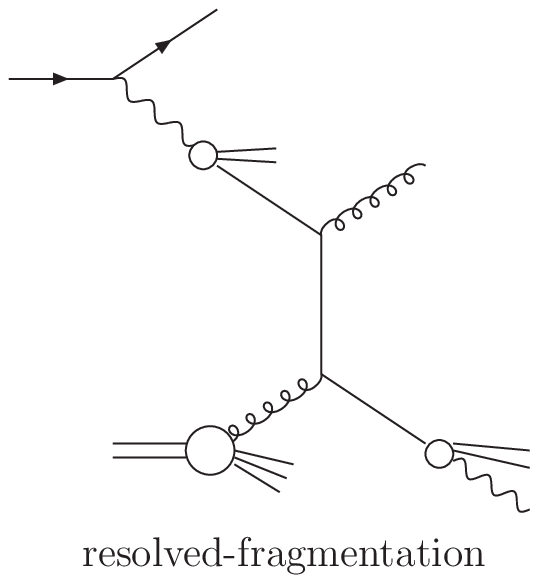,height=3.8cm}}
\end{picture}
\caption{Examples of contributing subprocesses at leading order to each of the four categories.}\label{fig:F1}
\end{figure}

The subprocesses contributing 
to the partonic reaction $ab \to c X$ can be divided into four 
categories,  
as depicted in Fig.~\ref{fig:F1}. 
The cases ``direct-direct" and ``resolved-direct" correspond to 
$c=\gamma$, so $D_{\gamma/c}(z,M_F)=\delta_{c\gamma}\delta(1-z)$ 
in (\ref{dsigma}), i.e. the prompt  
photon is produced directly in the hard subprocess and not from the 
fragmentation of a hard parton. 

The ``resolved" contributions are characterised by a resolved photon in the
initial state where a parton stemming from the  photon 
instead of the photon itself parti\-ci\-pates in the hard subprocess. 
In these cases, $F_{a/e}(x_e,M)$ is given by a convolution of the 
Weizs\"acker-Williams spectrum $f^e_{\gamma}(y)$ with the parton distributions 
in the photon:
\begin{eqnarray}
&&F_{a/e}(x_e,M)=\label{phpdf}\\
&&\int_0^1 dy \,dx_{\gamma}\,f^e_{\gamma}(y) \,
F_{a/\gamma}(x_{\gamma},M)\,\delta(x_{\gamma}y-x_e)\;.\nonumber
\end{eqnarray}
The cases with ``direct" attributed to the initial state photon 
correspond to  $a=\gamma$, so $F_{a/\gamma}=\delta(1-x_\gamma)$ and
 $F_{a/e}$ in eq.~(\ref{phpdf}) collapses to 
the Weizs\"acker-Williams spectrum. 

If additional jets are measured, eq.~(\ref{dsigma}) also contains a jet function, 
which defines the clustering of the final state partons other than the photon into jets.

\subsection*{Photon isolation}

In order to single out the prompt photon events from the background
of secondary photons produced by the decays of $\pi^0,\eta,\omega$ mesons,
isolation cuts have to be imposed on the photon signals in the experiment. 
A widely used isolation criterion is the following:
A photon is isolated if, inside a cone centered around the photon direction 
in the rapidity and azimuthal angle plane, the amount of hadronic transverse 
energy
$E_T^{had}$ deposited is smaller than some value $E_{T,\rm{max}}$\,: 
\begin{equation}\label{criterion}
\begin{array}{rcc} 
\mbox{for }\;\left(  \eta - \eta_{\gamma} \right)^{2} &+&  \left(  \phi - \phi_{\gamma} \right)^{2}  
 \leq   R,\\
E_T^{had} & \leq & E_{T,\rm{max}}\,.
\end{array}
\end{equation}
HERA experiments mostly used  
$E_{T,\rm{max}}=\epsilon \,p_T^{\gamma}$ with 
$\epsilon=0.1$ and $R$ = 1.
Isolation not only reduces the background from secondary photons, but also 
substantially  reduces the fragmentation components. 
It is important to note that the isolation parameters must be
carefully fixed in order to allow a comparison between data and perturbative 
QCD calculations. Indeed a part of the hadronic energy measured in the cone 
may come from the underlying event; therefore even the direct contribution 
can be cut by the isolation condition if the latter is too stringent. 

\subsection{Inclusive prompt photons}

\begin{figure}[htb]
\begin{picture}(150,130)(80,-40)
\put(50,-50){\epsfig{file=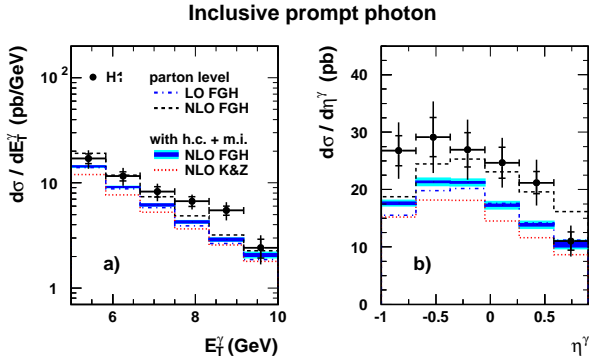,height=4.8cm}}
\end{picture}
\caption{H1 data\,\cite{Aktas:2004uv} compared to NLO
 QCD\,\cite{Krawczyk:2001tz,Fontannaz:2001ek}. 
 The NLO results are corrected for hadronisation and multiple interaction (h.c.+m.i.) 
 effects. The light blue (outer) error bands show the estimated uncertainties 
 on these corrections for the FGH result, the dark blue (inner) bands 
 show the scale uncertainty. The $\eta^\gamma$
 distribution is based on $E_{T,{\rm min}}^\gamma=5$\,GeV.}
\label{fig1}
\end{figure}

\begin{figure}[htb]
\begin{picture}(150,125)(50,-40)
\put(50,-50){\epsfig{file=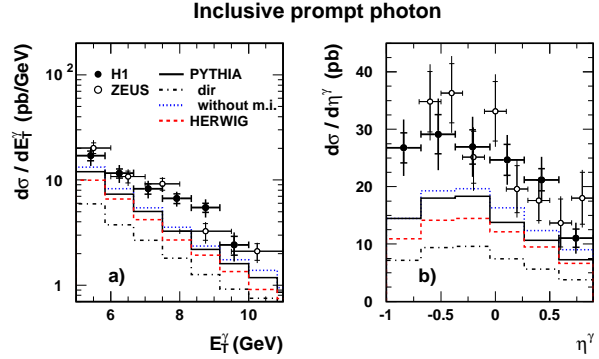,height=4.8cm}}
\end{picture}
\caption{H1 and ZEUS data on inclusive prompt photon production compared to {\small PYTHIA} and {\small HERWIG}. 
The blue histogram ``without multiple interactions (m.i.)" as well as
the ``direct only (dir)" histogram refer to the {\small PYTHIA} prediction. 
The figure is taken from ref.\,\cite{Aktas:2004uv}.}
\label{fig2}
\end{figure}

Inclusive prompt photons in photoproduction processes have been measured by 
ZEUS\,\cite{Breitweg:1999su} and H1 \cite{Aktas:2004uv}, and compared to 
{\small PYTHIA}\,\cite{Sjostrand:2000wi} and {\small HERWIG}\,\cite{Corcella:1999qn}
as well as NLO perturbative QCD calculations\,\cite{Krawczyk:2001tz,Fontannaz:2001ek}.
Interestingly, the data are above the NLO QCD prediction after 
corrections for hadronisation and multiple interactions (see Fig. \ref{fig1}).
This feature is seen in both, ZEUS and H1 data, and also persists 
in the comparison to  {\small PYTHIA}6.1 
and {\small HERWIG}6.1, as shown in Fig. \ref{fig2}.
The K\&Z\,\cite{Krawczyk:2001tz} curve being below the FGH\,\cite{Fontannaz:2001ek} 
curve in Fig.~\ref{fig1}
can be explained by the fact that the K\&Z calculation 
does not contain the  NLO corrections to the resolved part.

\subsection{Prompt photon + jet}

Prompt photon production in association with a jet offers more 
possibilities to probe the underlying parton dynamics, 
as it allows to define observables which give information about
the momentum fractions $x^\gamma,x^p$ the partons 
 are carrying with respect to the photon 
respectively proton they are originating from. 
The partonic $x^\gamma,x^p$  are not  observable, but 
  one can define the observables 
\begin{eqnarray}
x_{obs}^{\gamma}=
\frac{p_T^\gamma\,{\rm{e}}^{-\eta^\gamma}+p_T^{\rm{jet}}\,{\rm e}^{
-\eta^{\rm{jet}}}}{2E^{\gamma}}\;,\nonumber\\
x_{obs}^{p}=
\frac{p_T^\gamma\,{\rm{e}}^{\eta^\gamma}+p_T^{\rm{jet}}\,{\rm e}^{
\eta^{\rm{jet}}}}{2E^{p}}\;,
\label{xgam}
\end{eqnarray}
which, for direct photons in the final state, 
coincide with the partonic $x^\gamma,x^p$ at leading order.
Unique to photoproduction processes is the 
possibility to ``switch on/off" the resolved photon by suppressing/enhancing large
$x^\gamma$. 
As $x^\gamma=1$ corresponds to direct photons in the initial state, one can 
obtain e.g. resolved photon enriched data samples by placing a cut 
$x_{obs}^\gamma\leq 0.9$. Another possibility to enhance 
or suppress the resolved photon 
component is to place cuts on $p_T$ and rapidity.
From eq.~(\ref{xgam}) one can easily see that $x_{obs}^\gamma$ is 
small at low $p_T^{\gamma,\rm{jet}}$ values and large negative rapidities. 
Small $x^\gamma$-enriched data samples could be used to further 
constrain the parton distributions in the real photon, in particular 
the gluon distribution, as investigated e.g. in \cite{Fontannaz:2003yn}. 
Similarly, one can suppress the contribution from the resolved photon 
to probe the proton at small $x^p$ by direct 
$\gamma-p$ interactions\,\cite{Fontannaz:2003yn}. 

\begin{figure}[htb]
\begin{picture}(150,190)(70,60)
\put(85,38){\epsfig{file=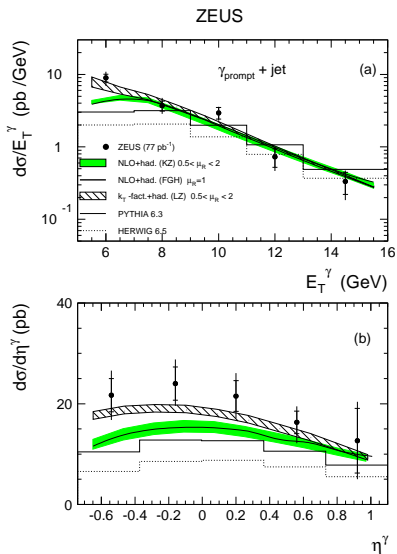,height=7.9cm}}
\end{picture}
\caption{ZEUS $\gamma$+jet data\,\cite{Chekanov:2006un} for $E_T^\gamma\geq 5$\,GeV
compared to different theory predictions (see text).}
\label{gjZeus5}
\end{figure}
As the hadronisation corrections are smaller for $\gamma$+jet than for dijet 
photoproduction\,\cite{Chekanov:2006un}, $\gamma$+jet final states in principle  
offer the possibility for highly accurate compa\-ri\-sons
of perturbative QCD predictions to the data, once the issues about 
photon isolation are well under control.
For example, a study of the effective transverse momentum $\langle k_T\rangle$ 
of partons in the proton has been made by ZEUS\,\cite{Chekanov:2001aq}. 
Comparing the shapes of normalised distributions for 
$\langle k_T\rangle$-sensitive ob\-ser\-vables to a NLO calculation, 
it was found that the data agree well with NLO QCD 
without extra intrinsic $\langle k_T\rangle$\,\cite{Fontannaz:2001nq}.
\begin{figure}[htb]
\begin{picture}(150,120)(70,50)
\put(65,39){\epsfig{file=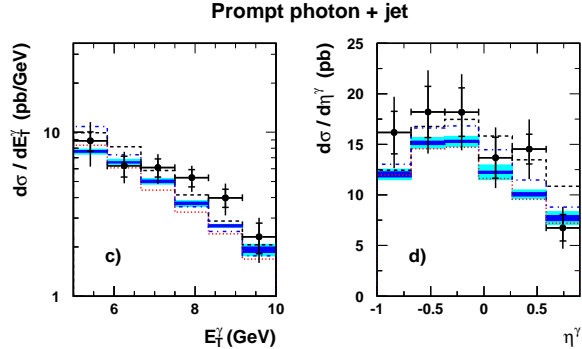,height=4.7cm}}
\end{picture}
\caption{H1 $\gamma$+jet data\,\cite{Aktas:2004uv} compared to NLO QCD predictions\,\cite{Krawczyk:2001tz,Fontannaz:2001nq}. 
The labelling of the histograms is the same as in Fig.\,\ref{fig1}.}
\label{gjH1}
\end{figure}

Detailed analyses for  $\gamma$+jet measurements 
in photoproduction have been carried out by both 
ZEUS\,\cite{Chekanov:2006un,Chekanov:2007cf} and 
H1\,\cite{Aktas:2004uv}, 
where the ZEUS collaboration has also compared with 
the $k_T$-factorisation approach of Lipatov and Zotov\,\cite{Lipatov:2005tz}.
Sample plots are shown in Figs.\,\ref{gjZeus5}
and \ref{gjH1}. 

The fact that in the last bin of Fig.\,\ref{gjH1}\,d) the data 
point drops below the parton level theory prediction (black dashed curve)
may be understood as an example where 
photon isolation in the partonic calculation acts  differently 
from photon isolation in the expe\-ri\-ment: In the large-$\eta^\gamma$ region, 
the contribution from resolved photons dominates, such that 
more remnants  from the resolved photon are present than in other rapidity domains. 
These remnants may deposit energy in the isolation cone which is above the 
$E_{T,\rm{max}}$ allowed by isolation, and thus the event is discarded, 
while this effect is certainly not fully captured by the partonic calculation.
This explanation is corroborated by the fact that the hadronisation corrections 
lower the parton level result substantially in the 
large-$\eta^\gamma$ region.

\begin{figure}[htb]
\begin{picture}(150,190)(70,60)
\put(90,40){\epsfig{file=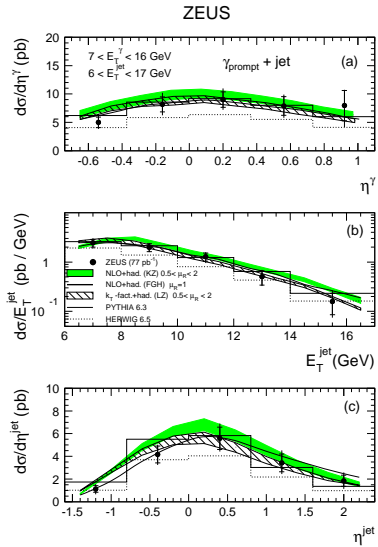,height=7.9cm}}
\end{picture}
\caption{ZEUS $\gamma$+jet data\,\cite{Chekanov:2006un} for $E_T^\gamma\geq 7$\,GeV
compared to different theory predictions.}
\label{gjZeus7}
\end{figure}

Interestingly, ZEUS investigated what happens if the 
minimum transverse energy of the prompt photon is increased 
to 7\,GeV, and found that in this case, 
the NLO calculations are in good agreement\,\cite{Chekanov:2006un} 
(see Fig.~\ref{gjZeus7}),  
which points to non-perturbative effects 
being the reason for the discrepancy.
In this light it would be interesting to reconsider the inclusive 
prompt photon data with increased $E_{T,{\rm min}}^\gamma$.


\section{Prompt photons in DIS}

Some time ago, the ZEUS collaboration 
performed a measurement\,\cite{Chekanov:2004wr}
of the inclusive prompt photon cross section in 
deeply inelastic scattering (DIS), 
where rather large discrepancies to the predictions of 
{\small PYTHIA}6.206\,\cite{Sjostrand:2001yu} and {\small HERWIG}6.1\,\cite{Corcella:1999qn}
were found. 
\begin{figure}[htb]
\begin{picture}(150,90)(50,50)
\put(44,40){\epsfig{file=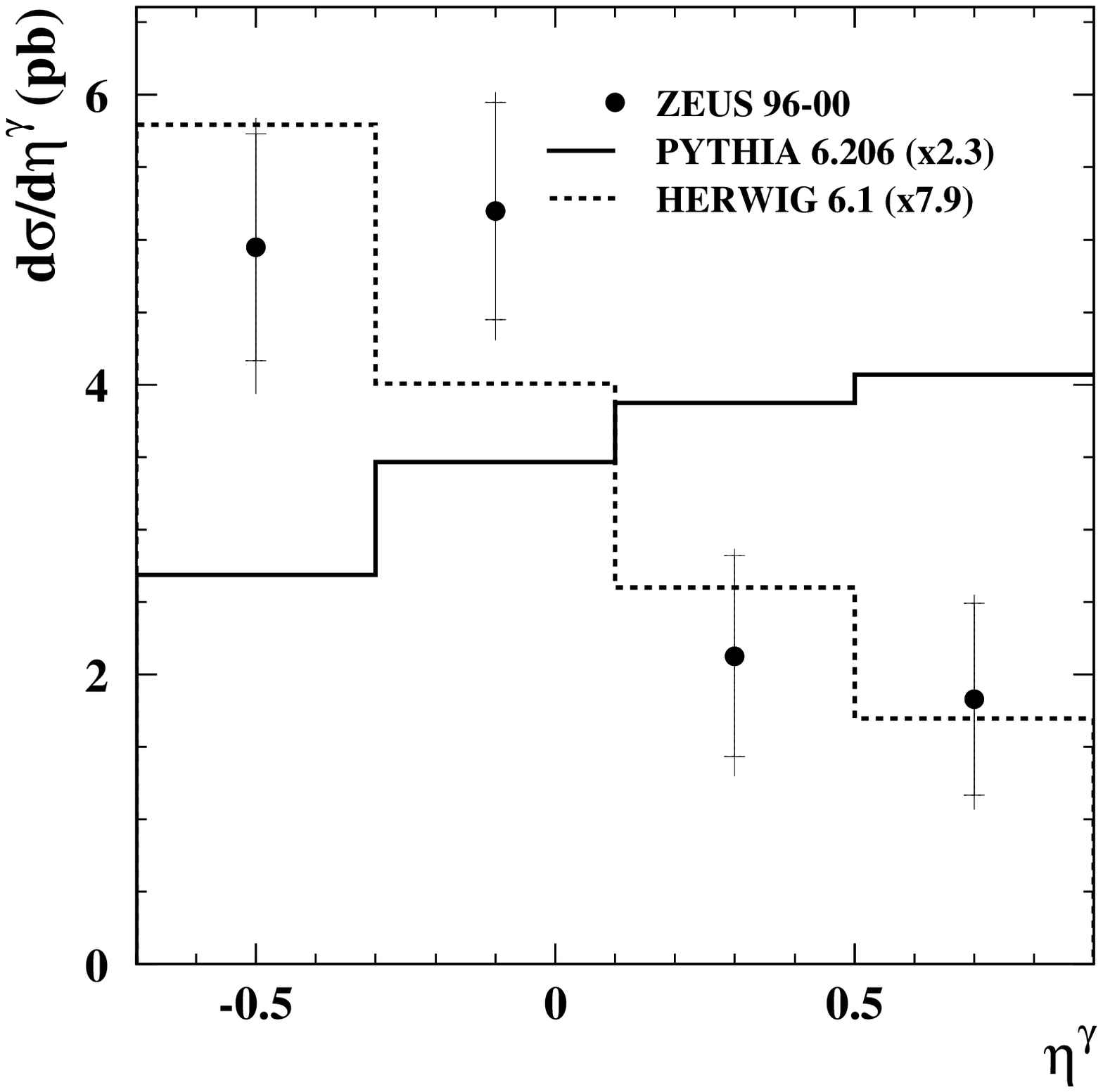,height=3.9cm}}
\put(153,135){\epsfig{file=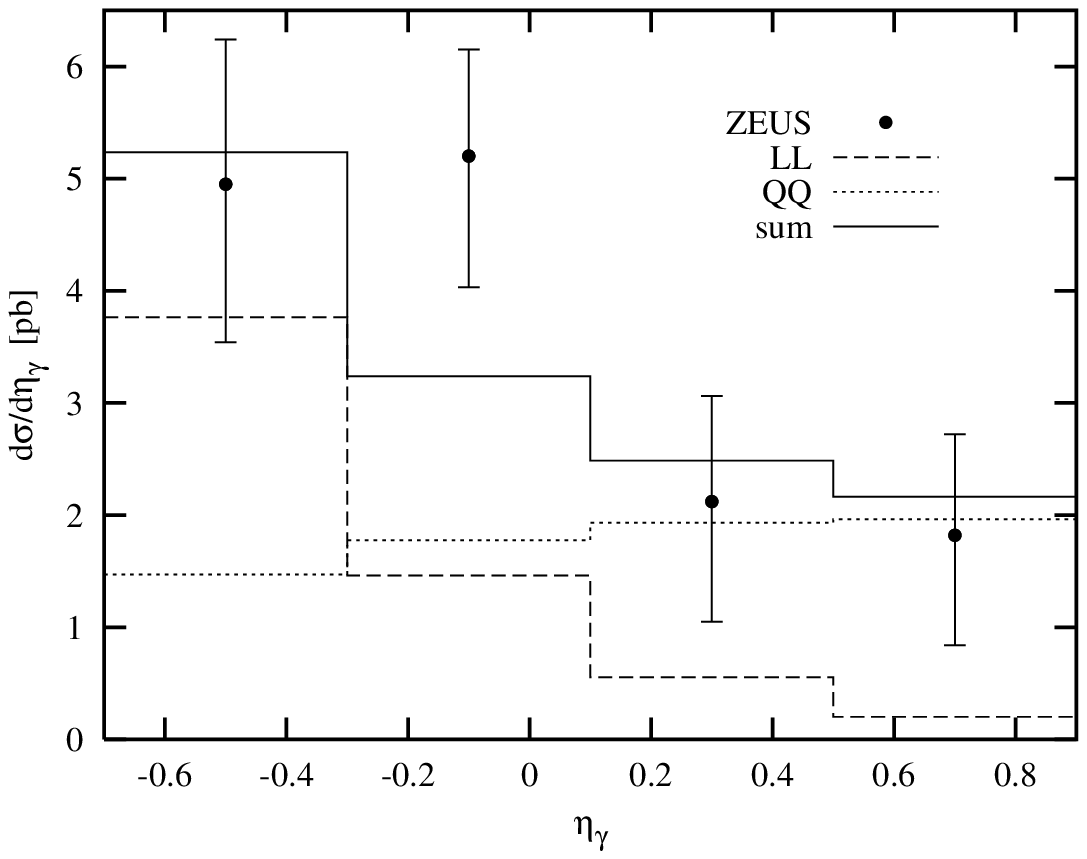,height=4.cm,angle=270}}
\end{picture}
\caption{Predictions from {\small PYTHIA} and {\small HERWIG} and predictions from a
parton level calculation\,\cite{GehrmannDeRidder:2006wz} 
compared to ZEUS data on inclusive prompt photon production in DIS\,\cite{Chekanov:2004wr}. LL denotes the contribution from 
photon radiation off leptons, QQ the one off quarks.}
\label{figDIS}
\end{figure}
This motivated a dedicated partonic calculation\,\cite{GehrmannDeRidder:2006wz} 
where both the radiation of photons off leptons and off quarks was  
taken into account. 
The result is in fair agreement with the experimental data,  
which shows the importance of both subprocesses,
as well as the inclusion of large-angle 
photon radiation\,\cite{GehrmannDeRidder:2006wz}. 
In particular, the shapes of the individual 
contributions in Fig.~\ref{figDIS} suggest that {\small PYTHIA} 
seems to underestimate the 
photon radiation off leptons. 
\begin{figure}[htb]
\begin{picture}(150,100)(50,50)
\put(63,35){\epsfig{file=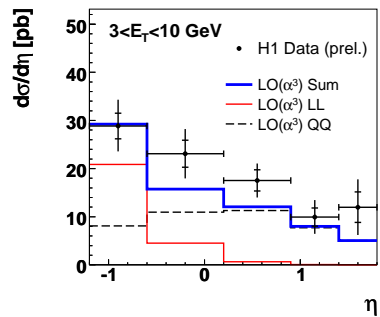,height=4.4cm}}
\end{picture}
\caption{Preliminary H1 data\,\cite{Schmitz:2006sc} compared to 
predictions from ref.\,\cite{GehrmannDeRidder:2006wz}. }
\label{H1DIS}
\end{figure}

Very recently, H1 also presented data on prompt photon production 
in DIS\,\cite{Schmitz:2006sc,MuellerSchmitz}, where the measurement range 
could be extended considerably as compared to previous measurements, 
thus increasing the cross section by almost an order of magnitude.
The $\eta^\gamma$ distribution, again compared to the calculation 
of ref.\,\cite{GehrmannDeRidder:2006wz}, 
is shown in Fig.\,\ref{H1DIS}.

Prompt photon plus jet production in DIS also has been measured 
by ZEUS\,\cite{Chekanov:2004wr} and H1\,\cite{MuellerSchmitz}, 
and compared to parton-level calculations. In this case, NLO predictions are 
available\,\cite{GehrmannDeRidder:2000ce,GehrmannDeRidder:1999yu,GehrmannDeRidder:1999wy}, 
which describe the shape of the $\eta^\gamma$ distribution reasonably well, but still 
underestimate the data\,\cite{Chekanov:2004wr,MuellerSchmitz}.

Based on the above mentioned partonic calculations, 
it is suggested in \cite{GehrmannDeRidder:2006vn} that
the photon fragmentation functions can be measured at HERA 
from $\gamma$+(0+1) -jet and $\gamma$+(1+1) -jet data samples in DIS.

\section{Prompt photons in hadronic collisions}

As already mentioned, prompt photon production in hadronic collisions 
is of particular importance these days for various reasons, e.g. photons play a 
major role in 
performing luminosity/calibration measurements at the LHC, 
and, at a later stage, measuring the gluon pdfs at the LHC.   
Further, (di-)photons are important in connection with the 
background to $H\to \gamma\gamma$ for $m_H\,\lsim \,140$\,GeV, 
pair-production of SUSY-particles decaying to lower mass states, 
radiative decays of excited states in various New Physics scenarios, etc.
Covering all these subjects is far beyond the scope of this short review, 
therefore we focus on calculations related to recent 
experimental measurements. 

It should be noted that at high energies, the electroweak corrections from
virtual weak boson exchange increase strongly. 
This fact motivated the calculation of the complete one-loop 
electroweak corrections to large-$p_T$ photon production 
in hadronic collisions\,\cite{Maina:2004rb,Kuhn:2005gv}.
Indeed it was found that at the LHC, where photon transverse momenta 
in the range of 2\,TeV are within reach, these corrections 
amount to up to -17\%\,\cite{Kuhn:2005gv}.

\subsection{Diphoton production in hadronic collisions}

Two publicly available parton level calculations which include 
various types of higher order corrections are available, 
{\small DIPHOX}\,\cite{Binoth:1999qq}  and ResBos\,\cite{Balazs:2007hr,Balazs:2006cc,Nadolsky:2007ba,Balazs:1997hv}.
ResBos has recently seen an important update\,\cite{Balazs:2007hr}, where 
 ${\cal O}(\alpha_s^3)$ corrections to $gg$-scattering
\cite{Bern:2001df,Bern:2002jx,Balazs:1999yf,deFlorian:1999tp}
have been included. In addition, it contains NNLL resummation of 
initial-state singularities at small $q_T$. Although a complete 
treatment of resummation would require joint initial- and final 
state resummation, which is quite difficult because of the interplay 
with photon isolation, the resummation done in \cite{Balazs:2007hr} 
certainly improves  the theo\-re\-tical prediction in the region 
sensitive to initial state multiple gluon emission. 
On the other hand, 
ResBos uses  an approximation for the fragmentation contributions
which is effectively leading order, 
while in {\small DIPHOX} the 
fragmentation contributions are included fully at next-to-leading order.

\begin{figure}[htb]
\begin{picture}(150,130)(50,50)
\put(45,38){\epsfig{file=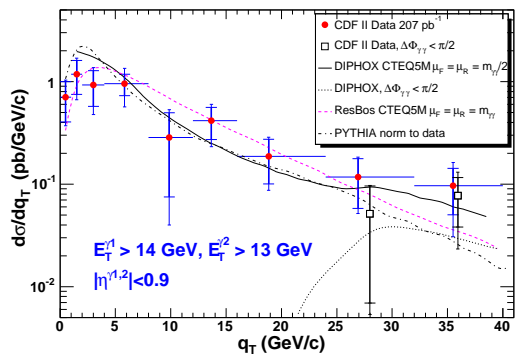,width=7.7cm}}
\end{picture}
\caption{CDF diphoton data compared to {\small DIPHOX}\,\cite{Binoth:1999qq},
 ResBos\,\cite{Balazs:1997hv} and {\small PYTHIA}6.216\,\cite{Sjostrand:2001yu}. 
The figure is taken from ref.\,\cite{Bandurin:2006bd}.}
\label{figdiphox}
\end{figure}
CDF has performed a recent measurement of diphoton 
production\,\cite{Acosta:2004sn}. The comparison to the data 
reflects the features of the different 
theoretical descriptions, as can be seen from Fig.~\ref{figdiphox}.
{\small DIPHOX} diverges at low $q_T$ because this region requires 
resummation of large logarithms. ResBos underestimates the tail 
because its fragmentation contribution is effectively at leading order. 
In particular, the shoulder seen in the data at $q_T\,\gsim\, 28$GeV 
can be understood as arising from an increase in phase space
for $\Delta\phi_{\gamma\gamma}<\pi/2$,    
which is particularly enhanced in the fragmentation contribution 
due to an interplay with isolation.
Detailed studies can be found in\,\cite{Binoth:2000zt}.
Note that Fig.~\ref{figdiphox} has been obtained with the ResBos 
version of ref.~\cite{Balazs:1997hv}, but the most recent version shows 
basically the 
same features with respect to this plot\,\cite{Balazs:2007hr}.

\subsection{Inclusive prompt photon production}

\begin{figure}[htb]
\begin{picture}(150,255)(50,50)
\put(55,36){\epsfig{file=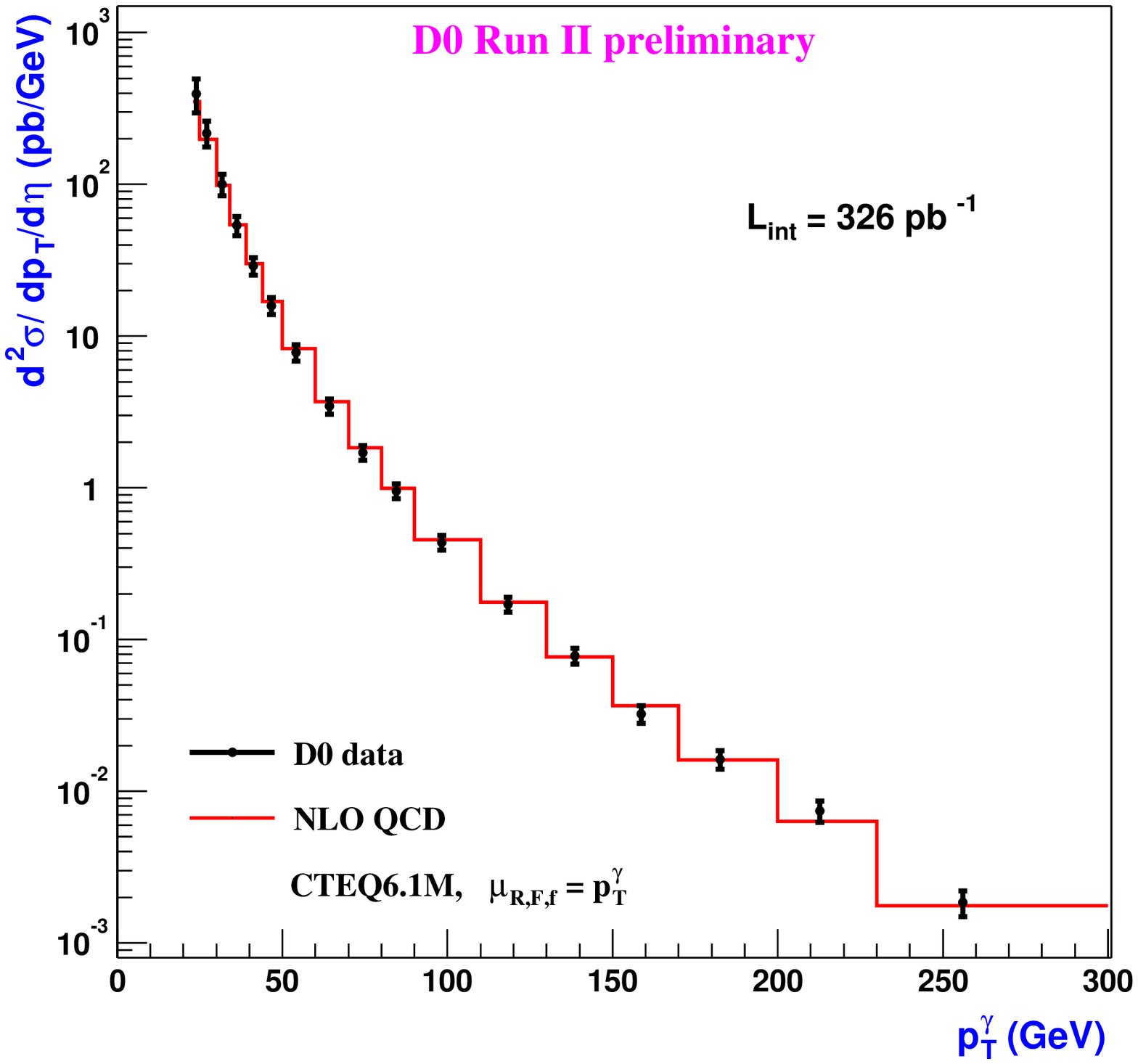,height=5.3cm}}
\put(55,180){\epsfig{file=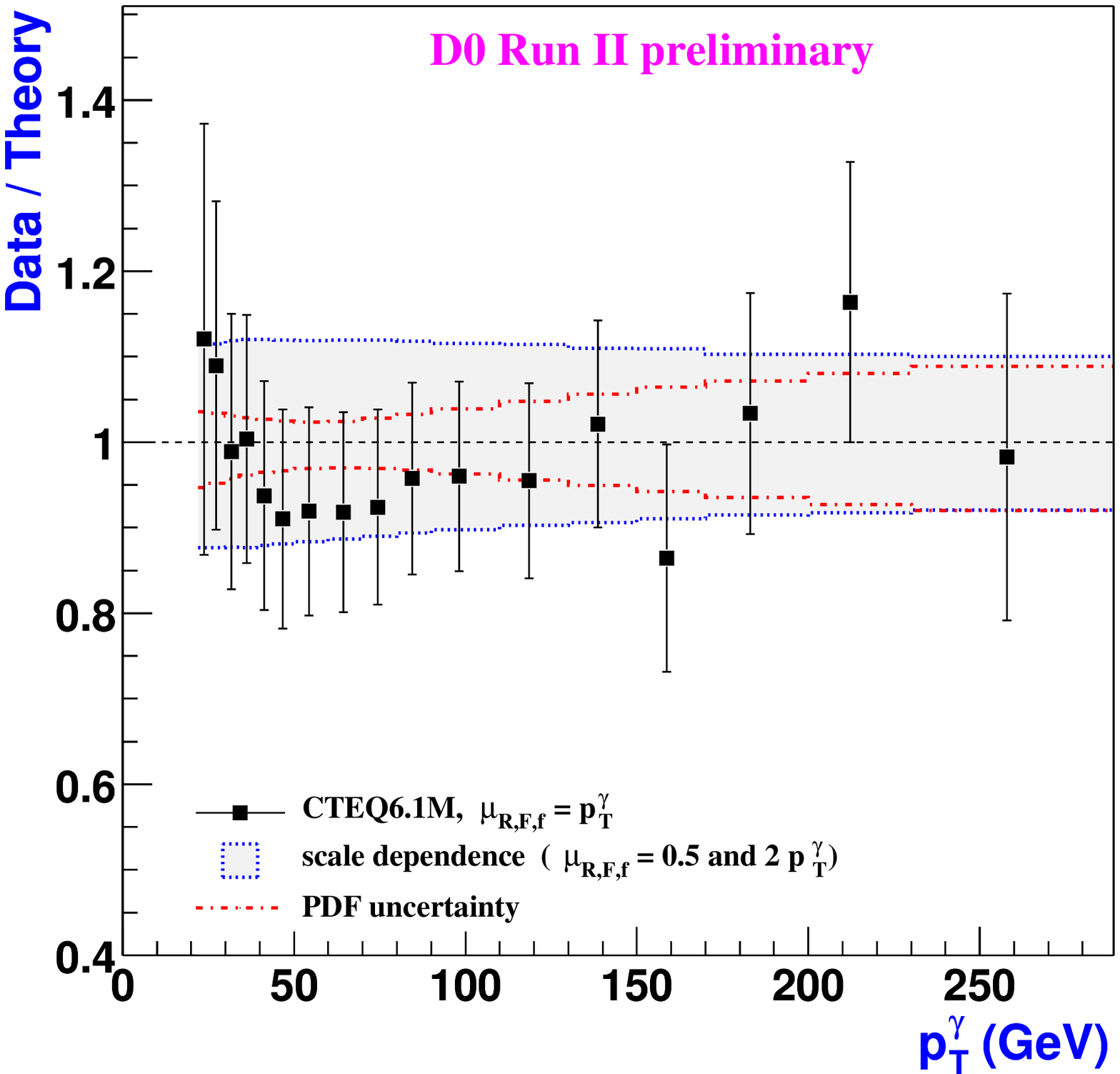,height=5.2cm}}
\end{picture}
\caption{D0 data\,\cite{Abazov:2005wc} on inclusive prompt photon production for 
$23\,\rm{GeV}\leq p_T^\gamma \leq 300$\,GeV, 
 $|\eta^\gamma|<0.9$ compared to {\small JETPHOX}\,\cite{Aurenche:2006vj}.
The figures are taken from \,\cite{SoldnerRembold:2005vz}.}
\label{figjetphox}
\end{figure}

D0 has measured the cross section for the inclusive production 
of isolated photons in the range $23<p_T^\gamma<300$\,GeV \cite{Abazov:2005wc}.
This extends previous measurements\,\cite{Acosta:2004bg,Acosta:2002ya,Abazov:2001af}
to significantly higher values of $p_T^\gamma$. 
In fact, the  $p_T^\gamma$ range is the widest ever tested.
As can be seen from Fig.\,\ref{figjetphox}, the data agree quite well 
with NLO QCD calculations  \cite{Aurenche:2006vj,Gordon:1994ut}.

\subsection{Prompt photon + jet production}

Measurements of $p\,\bar{p} \to \gamma +{\rm{jet}}+X$ for 
 30 GeV \,$\leq p_T^\gamma \leq 300 $\,GeV
are just being performed by D0\,\cite{AtramentovPH07}. 
The NLO partonic Monte Carlo program {\small JETPHOX}\,\cite{Aurenche:2006vj} is used to 
compare to theory at next-to-leading order.
The comparison is done separately for different regions in rapidity 
of the photon and the jet. The preliminary data  show a discrepancy to the  
theory prediction in some regions\,\cite{AtramentovPH07,Kumar:2007mf} 
which still needs to be understood, 
as it is not always at low $p_T^\gamma$, where it could be explained by 
non-perturbative effects.

Understanding prompt photon plus jet production in hadronic collisions 
is certainly very important in view of the LHC, where 
the process  
$g\,q \to q \gamma$ dominates in a wide kinematic range
and thus offers the possibility 
to constrain the gluon distribution in the
 proton\,\cite{Gupta:2007cy,Kumar:2003ue,Dissertori,ChekanovHERALHC07}. 
Further, $\gamma$+jet events can be used to set the absolute jet energy scale, 
as shown e.g. in dedicated CMS studies\,\cite{Konoplyanikov:2006ce,Adam:2005zf}.
It is also a good channel to study photon selection criteria in view of searches 
for New Physics\,\cite{Gupta:2007cy,ChekanovHERALHC07}.

\subsection{Prompt photon production at RHIC}

RHIC pp collisions at $\sqrt{s}=200$\,GeV\,\cite{Adler:2006yt,Isobe:2007ku}
cover the intermediate energy range between fixed target and 
Tevatron collider energies and are therefore of major importance 
to bridge a gap which allows a revisited interpretation 
of all of these data. 
Further, the PHENIX experiment at RHIC uses 
different  photon isolation methods\,\cite{Adler:2006yt}, 
which allows to study systematics
 related to isolation. 
An interesting possibility is also the measurement of photon-hadron azimuthal 
correlations as presented in \cite{Jin:2007by} and studied from the theory side in
\cite{Pietrycki:2007xr, PStheseproceedings}. 
\begin{figure}
\begin{picture}(150,130)(50,50)
\put(45,35){\epsfig{file=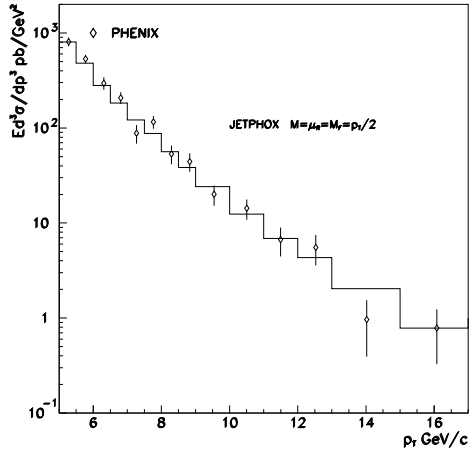,height=6.8cm}}
\end{picture}
\caption{Inclusive prompt photon data from PHENIX at $\sqrt{s}=200$\,GeV\,\cite{Adler:2006yt}
compared to {\small JETPHOX}.}
\label{figphenix}
\end{figure}

The  PHENIX prompt photon data have been compared to NLO QCD 
calculations by {\small JETPHOX}\,\cite{Aurenche:2006vj} and 
W.~Vogelsang {\it et al}\,\cite{Gordon:1994ut} and show very good agreement, 
as can be seen from Fig.~\ref{figphenix}.

\section{The overall picture}

Looking at Figs.\,\ref{figjetphox} and\,\ref{figphenix}, one may wonder 
why prompt photons in hadronic collisions had the bad reputation of 
exhibiting large
disagreements between NLO theory and data, partly in 
both normalisation and shape.
The discrepancies were mainly seen in prompt photon production 
on fixed targets, but also at Tevatron $p\bar{p}$ collider energies
(for a review, see e.g.\,\cite{Aurenche:2006vj,Aurenche:1998gv}).

It was suspected that there are large  effects from multiple soft gluon 
emission, and the necessity for large  intrinsic $\langle k_T\rangle$  
to account for soft gluon- and  nonperturbative effects 
was claimed.

\begin{figure}[htb]
\begin{picture}(150,155)(50,50)
\put(44,30){\epsfig{file=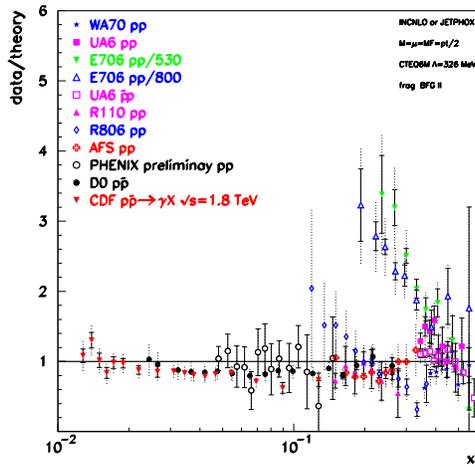,height=7.cm}}
\end{picture}
\caption{Collection of inclusive prompt photon data at different energies.
The figure is taken from ref.~\cite{Aurenche:2006vj}.}
\label{figalldata}
\end{figure}

As a consequence, various theory efforts to quantify these effects have been 
undertaken, e.g. threshold  
resummation for $x_T$=$2p_T/\sqrt{s}\to$ 1\,\cite{
deFlorian:2005wf,Bolzoni:2005xn,Kidonakis:2003bh,Sterman:2000pt,Catani:1999hs,Laenen:1998qw}
and joint resummation of  threshold and  recoil effects
\cite{Basu:2007nu,Sterman:2004yk,Kulesza:2002rh,Laenen:2000ij}.
It has been shown that the effect of resummation 
extends down to $x_T \,\,\gsim \,\,10^{-1}$, thus 
co\-ve\-ring the fixed target range, and that
scale dependences are considerably reduced by resummation.
Further, it turned out that, as to be expected, 
recoil effects in inclusive prompt photon production are 
relatively small, and that 
agreement with almost all prompt photon data
can be achieved. 
A detailed collection and analysis of prompt photon results 
at different energies has been performed in \cite{Aurenche:2006vj}, 
where Fig.\,\ref{figalldata} has been taken from. 
One can see that, with the new data from the Tevatron and 
from PHENIX, we cover a very wide range in $x_T$. 
More importantly, the picture emerges that, 
apart from the E706 data, 
the agreement with NLO theory is quite impressive.

In conclusion, we have learnt a lot (and still do\,!) 
from prompt photon production at   HERA.
Further, thanks to recent RHIC and 
 Tevatron measurements, 
the reputation of prompt photons in  hadronic collisions 
is rising again:
NLO QCD in general does a pretty good job where it is expected to do so.
Therefore we are looking forward to further exploiting the 
advantageous features of photons in future measurements and calculations.

\section*{Acknowledgements}
I would like to thank the organisers of the conference Photon\,2007 for the invitation,   
and M.~Fontannaz and J.-Ph.~Guillet for many fruitful discussions and 
collaborations centered around photons. I am also grateful to 
my experimental colleagues for interesting discussions 
about prompt photons, most of all P.~Bussey, S.~Chekanov, G.~Dissertori, J.~Gayler,  
J.~Huston, R.~Lemrani and M.~Werlen.
This research was supported 
 by the UK Science and Technology Facilities Council.


\begin{footnotesize}

\end{footnotesize}


\begin{thebibliography}{99}
\bibitem{url} Slides: \\ 
\verb$http://indico.cern.ch/materialDisplay.py?$ \\
\verb$contribId=36&sessionId=15$ \\
\verb$&materialId=slides&confId=3841$


   
\bibitem{Gluck:1992zx}
  M.~Gl\"uck, E.~Reya and A.~Vogt,
  Phys.\ Rev.\  D {\bf 48} (1993) 116
  [Erratum-ibid.\  D {\bf 51} (1995) 1427].
  
\bibitem{Bourhis:2000gs}
  L.~Bourhis, M.~Fontannaz, J.~P.~Guillet and M.~Werlen,
  Eur.\ Phys.\ J.\  C {\bf 19} (2001) 89
  [arXiv:hep-ph/0009101].


\bibitem{Krawczyk:2000mf}
  M.~Krawczyk, A.~Zembrzuski and M.~Staszel,
  Phys.\ Rept.\  {\bf 345} (2001) 265
  [arXiv:hep-ph/0011083].
  

 \bibitem{Aurenche:2005da}
  P.~Aurenche, M.~Fontannaz and J.~P.~Guillet,
  Eur.\ Phys.\ J.\  C {\bf 44} (2005) 395
  [arXiv:hep-ph/0503259].

 

\bibitem{Ajduk:2006gb}
  Z.~Ajduk, M.~Krawczyk and A.~K.~Wroblewski,
  Acta Phys.\ Polon.\  B {\bf 37} (2006) 543
  [Acta Phys.\ Polon.\  B {\bf 37} (2006) 1011].


\bibitem{Klasen:2002xb}
  M.~Klasen,
  Rev.\ Mod.\ Phys.\  {\bf 74} (2002) 1221
  [arXiv:hep-ph/0206169];\\
  M.~Klasen,
  arXiv:hep-ph/0702052.

\bibitem{WW}
C.F. Weizs\"acker, \zp{88}{34}{612};\\
 E.J. Williams, \pr{45}{34}{729}.
	
\bibitem{Frixione:1993yw}
  S.~Frixione, M.~L.~Mangano, P.~Nason and G.~Ridolfi,
  Phys.\ Lett.\  B {\bf 319} (1993) 339
  [arXiv:hep-ph/9310350].

\bibitem{Fontannaz:2003yn}
  M.~Fontannaz and G.~Heinrich,
  Eur.\ Phys.\ J.\  C {\bf 34} (2004) 191
  [arXiv:hep-ph/0312009].

\bibitem{Breitweg:1999su}
  J.~Breitweg {\it et al.}  [ZEUS Collaboration],
  Phys.\ Lett.\  B {\bf 472} (2000) 175
  [arXiv:hep-ex/9910045].

\bibitem{Aktas:2004uv}
  A.~Aktas {\it et al.}  [H1 Collaboration],
  Eur.\ Phys.\ J.\  C {\bf 38} (2005) 437
  [arXiv:hep-ex/0407018].

\bibitem{Sjostrand:2000wi}
  T.~Sjostrand, P.~Eden, C.~Friberg, L.~Lonnblad, G.~Miu, S.~Mrenna and E.~Norrbin,
  Comput.\ Phys.\ Commun.\  {\bf 135} (2001) 238
  [arXiv:hep-ph/0010017].

\bibitem{Corcella:1999qn}
  G.~Corcella {\it et al.},
  arXiv:hep-ph/9912396.

\bibitem{Krawczyk:2001tz}
  M.~Krawczyk and A.~Zembrzuski,
  Phys.\ Rev.\  D {\bf 64} (2001) 114017
  [arXiv:hep-ph/0105166].
  
\bibitem{Fontannaz:2001ek}
  M.~Fontannaz, J.~P.~Guillet and G.~Heinrich,
  Eur.\ Phys.\ J.\  C {\bf 21} (2001) 303
  [arXiv:hep-ph/0105121].

\bibitem{Chekanov:2006un}
  S.~Chekanov {\it et al.}  [ZEUS Collaboration],
  Eur.\ Phys.\ J.\  C {\bf 49} (2007) 511
  [arXiv:hep-ex/0608028].

\bibitem{Chekanov:2001aq}
  S.~Chekanov {\it et al.}  [ZEUS Collaboration],
  Phys.\ Lett.\  B {\bf 511} (2001) 19
  [arXiv:hep-ex/0104001].

\bibitem{Fontannaz:2001nq}
  M.~Fontannaz, J.~P.~Guillet and G.~Heinrich,
  Eur.\ Phys.\ J.\  C {\bf 22} (2001) 303
  [arXiv:hep-ph/0107262].

\bibitem{Chekanov:2007cf}
  S.~Chekanov  [ZEUS Collaboration], {\it talk given at DIS07}, 
  arXiv:0705.2700 [hep-ex].


\bibitem{Lipatov:2005tz}
  A.~V.~Lipatov and N.~P.~Zotov,
  Phys.\ Rev.\  D {\bf 72} (2005) 054002
  [arXiv:hep-ph/0506044].



\bibitem{Chekanov:2004wr}
  S.~Chekanov {\it et al.}  [ZEUS Collaboration],
  Phys.\ Lett.\  B {\bf 595} (2004) 86
  [arXiv:hep-ex/0402019].
  
\bibitem{Sjostrand:2001yu}
  T.~Sjostrand, L.~Lonnblad and S.~Mrenna,
  arXiv:hep-ph/0108264.
  
\bibitem{GehrmannDeRidder:2006wz}
  A.~Gehrmann-De Ridder, T.~Gehrmann and E.~Poulsen,
  Phys.\ Rev.\ Lett.\  {\bf 96} (2006) 132002
  [arXiv:hep-ph/0601073].

\bibitem{Schmitz:2006sc}
  C.~Schmitz  [H1 Collaboration],
  arXiv:hep-ex/0607093.

\bibitem{MuellerSchmitz}
  H1 Collaboration,
 H1prelim-07-033, 
{\it Abstract 370 submitted to the EPS2007 conference, Manchester, July 2007}.

\bibitem{GehrmannDeRidder:2000ce}
  A.~Gehrmann-De Ridder, G.~Kramer and H.~Spiesberger,
  Nucl.\ Phys.\  B {\bf 578} (2000) 326
  [arXiv:hep-ph/0003082].

\bibitem{GehrmannDeRidder:1999yu}
  A.~Gehrmann-De Ridder, G.~Kramer and H.~Spiesberger,
  Eur.\ Phys.\ J.\  C {\bf 11} (1999) 137
  [arXiv:hep-ph/9907511].

\bibitem{GehrmannDeRidder:1999wy}
  A.~Gehrmann-De Ridder, G.~Kramer and H.~Spiesberger,
  Phys.\ Lett.\  B {\bf 459} (1999) 271
  [arXiv:hep-ph/9903377].

\bibitem{GehrmannDeRidder:2006vn}
  A.~Gehrmann-De Ridder, T.~Gehrmann and E.~Poulsen,
  Eur.\ Phys.\ J.\  C {\bf 47} (2006) 395
  [arXiv:hep-ph/0604030].
  
  
\bibitem{Maina:2004rb}
  E.~Maina, S.~Moretti and D.~A.~Ross,
  Phys.\ Lett.\  B {\bf 593} (2004) 143
  [Erratum-ibid.\  B {\bf 614} (2005) 216]
  [arXiv:hep-ph/0403050].
  
\bibitem{Kuhn:2005gv}
  J.~H.~K\"uhn, A.~Kulesza, S.~Pozzorini and M.~Schulze,
  JHEP {\bf 0603} (2006) 059
  [arXiv:hep-ph/0508253].
 
  

\bibitem{Binoth:1999qq}
  T.~Binoth, J.~P.~Guillet, E.~Pilon and M.~Werlen,
  Eur.\ Phys.\ J.\  C {\bf 16} (2000) 311
  [arXiv:hep-ph/9911340].


\bibitem{Balazs:2007hr}
  C.~Balazs, E.~L.~Berger, P.~M.~Nadolsky and C.~P.~Yuan,
  Phys.\ Rev.\  D {\bf 76} (2007) 013009
  [arXiv:0704.0001 [hep-ph]].
  
\bibitem{Balazs:2006cc}
  C.~Balazs, E.~L.~Berger, P.~Nadolsky and C.~P.~Yuan,
  Phys.\ Lett.\  B {\bf 637} (2006) 235
  [arXiv:hep-ph/0603037].
  
\bibitem{Nadolsky:2007ba}
  P.~Nadolsky, C.~Balazs, E.~L.~Berger and C.~P.~Yuan,
  Phys.\ Rev.\  D {\bf 76} (2007) 013008
  [arXiv:hep-ph/0702003].
  
  
\bibitem{Balazs:1997hv}
  C.~Balazs, E.~L.~Berger, S.~Mrenna and C.~P.~Yuan,
  Phys.\ Rev.\  D {\bf 57} (1998) 6934
  [arXiv:hep-ph/9712471].
  
  
  
\bibitem{Bern:2001df}
  Z.~Bern, A.~De Freitas and L.~J.~Dixon,
  JHEP {\bf 0109} (2001) 037
  [arXiv:hep-ph/0109078].
  
\bibitem{Bern:2002jx}
  Z.~Bern, L.~J.~Dixon and C.~Schmidt,
  Phys.\ Rev.\  D {\bf 66} (2002) 074018
  [arXiv:hep-ph/0206194].
  
\bibitem{Balazs:1999yf}
  C.~Balazs, P.~Nadolsky, C.~Schmidt and C.~P.~Yuan,
  Phys.\ Lett.\  B {\bf 489} (2000) 157
  [arXiv:hep-ph/9905551].
  
  
\bibitem{deFlorian:1999tp}
  D.~de Florian and Z.~Kunszt,
  Phys.\ Lett.\  B {\bf 460} (1999) 184
  [arXiv:hep-ph/9905283].
  
 
\bibitem{Bandurin:2006bd}
  D.~Bandurin,
  AIP Conf.\ Proc.\  {\bf 870} (2006) 416.
  
\bibitem{Binoth:2000zt}
  T.~Binoth, J.~P.~Guillet, E.~Pilon and M.~Werlen,
  Phys.\ Rev.\  D {\bf 63} (2001) 114016
  [arXiv:hep-ph/0012191].

\bibitem{Acosta:2004sn}
  D.~E.~Acosta {\it et al.}  [CDF Collaboration],
  Phys.\ Rev.\ Lett.\  {\bf 95} (2005) 022003
  [arXiv:hep-ex/0412050].


\bibitem{Abazov:2005wc}
  V.~M.~Abazov {\it et al.}  [D0 Collaboration],
  Phys.\ Lett.\  B {\bf 639} (2006) 151
  [arXiv:hep-ex/0511054].

\bibitem{Acosta:2004bg}
  D.~E.~Acosta {\it et al.}  [CDF Collaboration],
  Phys.\ Rev.\  D {\bf 70} (2004) 074008
  [arXiv:hep-ex/0404022].

\bibitem{Acosta:2002ya}
  D.~E.~Acosta {\it et al.}  [CDF Collaboration],
  Phys.\ Rev.\  D {\bf 65} (2002) 112003
  [arXiv:hep-ex/0201004].
 
\bibitem{Abazov:2001af}
  V.~M.~Abazov {\it et al.}  [D0 Collaboration],
  Phys.\ Rev.\ Lett.\  {\bf 87} (2001) 251805
  [arXiv:hep-ex/0106026].


 \bibitem{Aurenche:2006vj}
  P.~Aurenche, M.~Fontannaz, J.~P.~Guillet, E.~Pilon and M.~Werlen,
  Phys.\ Rev.\  D {\bf 73} (2006) 094007
  [arXiv:hep-ph/0602133].\\
  For the program JETPHOX, see {\tt http://wwwlapp.in2p3.fr/lapth/PHOX\_FAMILY/}\\
  {\tt main.html}.

\bibitem{Gordon:1994ut}
  L.~E.~Gordon and W.~Vogelsang,
  Phys.\ Rev.\  D {\bf 50} (1994) 1901.
  

\bibitem{SoldnerRembold:2005vz}
  S.~S\"oldner-Rembold  [D0 Collaboration],
  Acta Phys.\ Polon.\  B {\bf 37} (2006) 733
  [arXiv:hep-ex/0511051].
  
 \bibitem{AtramentovPH07}
 O.~Atramentov, {\it these proceedings}. 

\bibitem{Kumar:2007mf}
  A.~Kumar  [D0 Collaboration],
  arXiv:0710.0415 [hep-ex].
 
\bibitem{Gupta:2007cy}
  P.~Gupta, B.~C.~Choudhary, S.~Chatterji, S.~Bhattacharya and R.~K.~Shivpuri,
  arXiv:0705.2740 [hep-ex].
  
\bibitem{Kumar:2003ue}
  A.~Kumar, M.~Kumar Jha, B.~Mitra Sodermark, A.~Bhardwaj, K.~Ranjan and R.~K.~Shivpuri,
  Phys.\ Rev.\  D {\bf 67} (2003) 014016.


\bibitem{Dissertori}  
G.~Dissertori, 
{\it Talk given at the Les Houches 2007 workshop on 
``Physics at TeV Colliders"}, 
to appear in the proceedings.

\bibitem{ChekanovHERALHC07}  
S.~Chekanov, {\it Talk given at the HERA-LHC workshop, DESY, October 2007,}
to apppear in the proceedings.

\bibitem{Konoplyanikov:2006ce}
  V.~Konoplyanikov, O.~Kodolova and A.~Ulyanov,
  CERN-CMS-NOTE-2006-042.
  
\bibitem{Adam:2005zf}
  W.~Adam {\it et al.}  [CMS Trigger and Data Acquisition Group],
  Eur.\ Phys.\ J.\  C {\bf 46} (2006) 605
  [arXiv:hep-ex/0512077].
  
  



\bibitem{Adler:2006yt}
  S.~S.~Adler {\it et al.}  [PHENIX Collaboration],
  Phys.\ Rev.\ Lett.\  {\bf 98} (2007) 012002
  [arXiv:hep-ex/0609031].
  
\bibitem{Isobe:2007ku}
  T.~Isobe  [PHENIX Collaboration],
  J.\ Phys.\ G {\bf 34} (2007) S1015
  [arXiv:nucl-ex/0701040].
  
\bibitem{Jin:2007by}
  J.~Jin  [PHENIX Collaboration],
  J.\ Phys.\ G {\bf 34} (2007) S813
  [arXiv:0705.0842 [nucl-ex]].

\bibitem{Pietrycki:2007xr}
  T.~Pietrycki and A.~Szczurek,
  Phys.\ Rev.\  D {\bf 76} (2007) 034003
  [arXiv:0704.2158 [hep-ph]].
  
  \bibitem{PStheseproceedings}
  A.~Szczurek, {\it these proceedings}.

\bibitem{Aurenche:1998gv}
  P.~Aurenche, M.~Fontannaz, J.~P.~Guillet, B.~A.~Kniehl, E.~Pilon and M.~Werlen,
  Eur.\ Phys.\ J.\  C {\bf 9} (1999) 107
  [arXiv:hep-ph/9811382].


  
\bibitem{deFlorian:2005wf}
  D.~de Florian and W.~Vogelsang,
  Phys.\ Rev.\  D {\bf 72} (2005) 014014
  [arXiv:hep-ph/0506150].

 
\bibitem{Bolzoni:2005xn}
  P.~Bolzoni, S.~Forte and G.~Ridolfi,
  Nucl.\ Phys.\  B {\bf 731} (2005) 85
  [arXiv:hep-ph/0504115].
  
\bibitem{Kidonakis:2003bh}
  N.~Kidonakis and J.~F.~Owens,
  Int.\ J.\ Mod.\ Phys.\  A {\bf 19} (2004) 149
  [arXiv:hep-ph/0307352];
  Phys.\ Rev.\  D {\bf 61} (2000) 094004
  [arXiv:hep-ph/9912388].
  
\bibitem{Sterman:2000pt}
  G.~Sterman and W.~Vogelsang,
  JHEP {\bf 0102} (2001) 016
  [arXiv:hep-ph/0011289].


\bibitem{Catani:1999hs}
  S.~Catani, M.~L.~Mangano, P.~Nason, C.~Oleari and W.~Vogelsang,
  JHEP {\bf 9903} (1999) 025
  [arXiv:hep-ph/9903436].
  
    
\bibitem{Laenen:1998qw}
  E.~Laenen, G.~Oderda and G.~Sterman,
  Phys.\ Lett.\  B {\bf 438} (1998) 173
  [arXiv:hep-ph/9806467].
  
  
  
\bibitem{Basu:2007nu}
  R.~Basu, E.~Laenen, A.~Misra and P.~Motylinski,
  Phys.\ Rev.\  D {\bf 76} (2007) 014010
  [arXiv:0704.3180 [hep-ph]].

\bibitem{Sterman:2004yk}
  G.~Sterman and W.~Vogelsang,
  Phys.\ Rev.\  D {\bf 71} (2005) 014013
  [arXiv:hep-ph/0409234].

\bibitem{Kulesza:2002rh}
  A.~Kulesza, G.~Sterman and W.~Vogelsang,
  Phys.\ Rev.\  D {\bf 66} (2002) 014011
  [arXiv:hep-ph/0202251].

\bibitem{Laenen:2000ij}
  E.~Laenen, G.~Sterman and W.~Vogelsang,
  Phys.\ Rev.\  D {\bf 63} (2001) 114018
  [arXiv:hep-ph/0010080].
 
  
 \end{thebibliography}
\end{document}